\documentstyle[11pt,paspconf,epsf]{article}

\def\kms{\mbox{$\rm km\;s^{-1}$}}

\begin{document}

\title{Kinematics of Gas and Stars in Early-type Disk Galaxies}
\author{J.C. Vega Beltr\'an, E. Pignatelli}
\affil{Osservatorio Astronomico di Padova, Italy}
\author{W.W. Zeilinger}
\affil{Insitut f\"ur Astronomie, Universit\"at Wien, Austria}
\author{A. Pizzella}
\affil{European Southern Observatory, Santiago, Chile}
\author{E.M. Corsini, F. Bertola}
\affil{Dipartimento di Astronomia, Universit\`a di Padova, Italy}
\author{J.E. Beckman}
\affil{Instituto Astrof\'{\i}sico de Canarias, Spain}

\rule{0cm}{1cm}


We present a comparison between ionized gas and stellar
kinematics for a sample of 5 early-to-intermediate disk galaxies
(see Table~1).  We measured the major axis $V$ and $\sigma$
radial profiles for both gas and stars, and the $h_3$ and $h_4$ radial
profiles of the stars. We also derived from the $R$-band surface
photometry of each galaxy the light contribution of their bulges
and disks.
 
In general, in late-type spirals both the
gas and the stars are found to be in virtually circular motions and
are characterized by a constant low value of the velocity
dispersion.  In some early-type disk galaxies (e.g. NGC~2775) the gas shows a slowly increasing
rotation velocity and a velocity dispersion peaked towards the center
over an extended radial range implying that even for the gaseous
component the rotation velocity falls short of the circular velocity
predicted by the kinematic model as found for S0's by Bertola et
al. 1995.

In order to investigate these differences in the velocity fields 
we adopted the self-consistent dynamical model by Pignatelli \&
Galletta (1998), which takes into account the asymmetric
drift  effects, the projection effects along the
line-of-sight and the non-Gaussian shape of the line
profiles due to the presence of different components with distinct
dynamical behavior. We find for the stellar component a sizeable
asymmetric drift effect in the inner regions of all the sample
galaxies, as it results by comparing their stellar rotation curves
with the circular velocity predicted by the models.

On the other hand, the gas component shows a very wide range of
kinematical properties.  Two of the sample galaxies (NGC~5064 and
NGC~7782) show almost circular motions even very near the center; two
present slowly rising rotation curves (NGC~772 and NGC~3898) and one
(NGC~980) even appears to have a gas rotation velocity {\em lower} than
that of the stellar component.

\begin{center}
{\bf Table 1:} Parameters of the Sample Galaxies \vspace*{0.2cm} \\
\begin{footnotesize}
\begin{tabular}{lllrccccr}
\hline
\noalign{\smallskip}
\multicolumn{1}{c}{object} &
\multicolumn{1}{c}{type} &
\multicolumn{1}{c}{$B_T$} &
\multicolumn{1}{c}{P.A.} &
\multicolumn{1}{c}{$i$} &
\multicolumn{1}{c}{$V_{\odot}$} &
\multicolumn{1}{c}{$V_0$} &
\multicolumn{1}{c}{$D$} &
\multicolumn{1}{c}{scale} 
						 \\
\multicolumn{1}{c}{\footnotesize [name]} &
\multicolumn{1}{c}{\footnotesize [RC3]} &
\multicolumn{1}{c}{\footnotesize [mag]} &
\multicolumn{1}{c}{\footnotesize [\deg]} &
\multicolumn{1}{c}{\footnotesize [\deg]} &
\multicolumn{1}{c}{\footnotesize [\kms]} &
\multicolumn{1}{c}{\footnotesize [\kms]} &
\multicolumn{1}{c}{\footnotesize [Mpc]} &
\multicolumn{1}{c}{\footnotesize [pc$''^{-1}$]} 
						\\
\multicolumn{1}{c}{({1})} &
\multicolumn{1}{c}{({2})} &
\multicolumn{1}{c}{({3})} &
\multicolumn{1}{c}{({4})} &
\multicolumn{1}{c}{({5})} &
\multicolumn{1}{c}{({6})} &
\multicolumn{1}{c}{({7})} &
\multicolumn{1}{c}{({8})} &
\multicolumn{1}{c}{({9})} 
						\\

\noalign{\smallskip}

\hline
\noalign{\smallskip}
NGC  772 &.SAS3.. & 11.09 & 130 &  53  &$2470\pm 10$&$2618$& 34.7 &  168   \\
NGC  980 &.L..... & 13.16 & 110 &  62  &$5765\pm 10$&$5936$& 79.1 &  384   \\
NGC 3898 &.SAS2.. & 11.60 & 107 &  46  &$1184\pm 10$&$1283$& 17.1 &   83   \\
NGC 5064 &PSA.2*. & 12.69 &  38 &  65  &$2980\pm 10$&$2750$& 36.7 &  178   \\
NGC 7782 &.SAS3.. & 13.08 & 175 &  62  &$5430\pm 10$&$5611$& 74.8 &  363   \\
\hline
\end{tabular} 

\begin{minipage}{12.5cm}
{NOTES} -- 
Col.({2}): classification from RC3 (de Vaucouleurs et al. 1991).
Col.({3}): total observed blue magnitude from RC3 except for NGC~980 from LEDA database.
Col.({4}): position angle of the major axis (RC3). 
Col.({5}): inclination from NGB (Tully 1988) except for NGC~980 and NGC~7782 from RC3
(inferred from R$_{25}$).
Col.({6}): heliocentric velocity of the galaxy derived as center of symmetry
of the gas rotation curve.
Col.({7}): systemic velocity derived from $V_\odot$ corrected for the motion 
of the Sun with respect of the Local Group by $\Delta V=300\cos{b}\sin{l}$.
Col.({8}): distance obtained as $V_0/H_0$ with $H_0=75$ \kms\ Mpc$^{-1}$.  
\end{minipage}
\end{footnotesize}
\end{center}

The galaxy sample is not wide enough to draw general conclusions.
 Anyway we found a possible correlation between the presence of
slowly-rising gas rotation curves and the ratio of the bulge/disk
half luminosity radii, while there is no obvious correlation with the key
parameter represented by the morphological classification,  namely the bulge/disk
luminosity ratio.  Systems with a diffuse dynamically hot component
(bulge or lens) with  a scale length comparable to that of the
disk are characterized by slowly-rising gas rotation curves. On the
other hand, in systems with a small bulge the gas follows almost
circular motions, regardless of the luminosity of the bulge itself.
We noticed a similar behaviour also in the gas and stellar
kinematics of the two early-type spiral galaxies modeled by Corsini et
al. (1998).

\end{document}